%% file: TempGrad.tex
\documentclass[12pt]{revtex4-2}
\expandafter\let\csname equation*\endcsname\relax

\expandafter\let\csname endequation*\endcsname\relax
\usepackage{color}
\usepackage{braket}
\usepackage{amsmath}

\usepackage{epsfig}
\usepackage[outdir=./figs/]{epstopdf}
\usepackage[usenames,dvipsnames,table]{xcolor}
\usepackage[colorlinks=true]{hyperref}
\usepackage{cleveref}
\usepackage[normalem]{ulem}
\newcommand{\beq}{\begin{equation}}
\newcommand{\eeq}{\end{equation}}
\newcommand{\bea}{\begin{align}}
\newcommand{\eea}{\end{align}}
\newcommand{\om}{\omega}
\newcommand{\Om}{\Omega}

\newcommand{\ga}{\gamma}

\newcommand{\D}{\Delta}

\newcommand{\h}{\hbar}
\newcommand{\ajd}{\hat a_j^{\dagger}}
\newcommand{\aj}{\hat a_j}
\newcommand{\qj}{\hat q_j}
\newcommand{\pj}{\hat p_j}
\newcommand{\ham}{\hat H}

\newcommand{\ai}{\hat a_1}

\newcommand{\aii}{\hat a_2}
\newcommand{\ain}{\hat a^{\text{in}}}
\newcommand{\aind}{\hat a^{\text{in}\dagger}}
\newcommand{\uop}{\hat u}

\newcommand{\fl}[1]{\delta #1}

\newcommand{\av}[1]{\langle #1 \rangle}

\newcommand\minus{\text{-}}

\begin{document}

\title[]{Temperature gradient and asymmetric steady state correlations in dissipatively coupled cascaded optomechanical systems}

\author{Claudio Pellitteri $^{1}$, G. Massimo Palma $^{1,2}$ and Salvatore Lorenzo$^{1}$}

\address{$^1$ Università degli Studi di Palermo, Dipartimento di Fisica e Chimica - Emilio Segrè, via Archirafi 36, I-90123 Palermo, Italy\\
$^2$ NEST, Istituto Nanoscienze-CNR, Piazza S. Silvestro 12, 56127 Pisa, Italy}
\vspace{10pt}
%\begin{indented}
%\item[]
%\end{indented}

\begin{abstract}
The interaction between a light mode and a mechanical oscillator via radiation pressure in optomechanical systems is an excellent platform for a multitude of applications in quantum technologies. In this work we study the dynamics of a pair of optomechanical systems interacting  dissipatively with a wave guide in a unidirectional way. Focusing on the regime where the cavity modes can be adiabatically eliminated we derive an effective coupling between the two mechanical modes and we explore both classical and quantum correlations established between the modes in both in the transient and in the stationary regime, highlighting their asymmmetrical nature due to the unidirectional coupling, and we find that a constant amount of steady correlations can exist at long times. Furthermore we show that this unidirectional coupling establishes a temperature gradient between the mirrors, depending on the frequencies' detuning. We additionally analyze the power spectrum of the output guide field and we show how, thanks to the chiral coupling, from such spectrum  it is possible to reconstruct the spectra of each single mirror.
\end{abstract}
\maketitle
%\maketitle

\section{Introduction}

Optomechanical systems, with light modes interacting with massive mechanical oscillators, have attracted a considerable interest for their possible application in quantum technologies \cite{aspelmeyer_quantum_2010,bowen_quantum_2020}.
Depending on the working point, the optomechanical interaction can be used to cool the mechanical mode near to its ground state \cite{genes_ground-state_2008,gigan_self-cooling_2006,marquardt_quantum_2007,marquardt_quantum_2008,yang_ground-state_2019,vinh2021} (a technique applied also to levitating nanospheres \cite{delic_cooling_2020}), to generate squeezing \cite{lecocq_quantum_2015,pirkkalainen_squeezing_2015,squeeze_2022} or to create entanglement between optical and mechanical modes \cite{palomaki_entangling_2013,riedinger_non-classical_2016,gut_stationary_2020}. These configurations can be mixed in an appropriate way in order to generate purely quantum states of the mechanical oscillators (e.g. generation of single phonon states \cite{Hong}). 

A natural extension of the standard  single mode - single mirror oscillators setups consists of several coupled modes.  We can distinguish two major and distinct setups. The first one is called {\it Multimode optomechanical system} \cite{bhattacharya_multiple_2008,xuereb_reconfigurable_2014,xuereb_strong_2012,weaver_coherent_2017-1,piergentili_two-membrane_2021}, and consists of  several mechanical oscillator interacting with the same cavity. In {\it Optomechanical array} instead, each mechanical oscillator interacts locally with its own cavity-mode but an effective coupling between neighbouring mirrors is implemented by photons and/or phonons tunneling \cite{peano_topological_2015,gan_solitons_2016,heinrich_collective_2011}.

In this work we propose the largely unexplored setup in which the cavity modes are dissipatively coupled via a unidirectional waveguide \cite{gil-santos_light-mediated_2017} in a cascaded configuration \cite{pichler_quantum_2015,giovannetti_master_2012-3,giovannetti_master_2012-2,ramos_quantum_2014,cusumano_interferometric_2018-1}. This arrangement induces a non-reciprocal interaction, at first between the cavities and then between the mechanical oscillators \cite{karg_remote_2019}. While previous studies have explored similar setups,  none of them has specifically addressed pure unidirectional coupling \cite{Farace2014}.
 For instance, in \cite{li_long-distance_2016,Li_Jin}, the authors study the synchronization between two resonators driven by a blue detuned laser, which leads to a self-sustained oscillatory dynamics and in \cite{xuereb_routing} the possibility of creating non-reciprocal devices that control the flow of thermal noise towards or away from specific quantum devices in a network has been explored. However, in spite of the above, the full potential of cascaded coupling between a pair of optomechanical systems remains largely unexplored, especially when it comes to the effective coupling between the two mechanical modes.
To fill this gap, we derive the equations describing the effective coupling between the two mirrors by adiabatically eliminating the cavity modes. We then characterise the correlations established between the mirrors in terms of mutual information and quantum discord, the latter being the best quantifier when one is interested in asymmetries between the two mirrors. We show that asymmetric non-zero correlations persists even in the steady state. Additionally, we explore the consequences of the asymmetry in the coupling, arising from the unidirectionality, and its implications for establishing a temperature gradient between the two modes. Notably, this temperature gradient vanishes in the case of perfectly symmetrical bidirectional coupling.

This work is organized as follow: In \cref{sec2} we present our model, we introduce its Hamiltonian. By employing Langevin equations we characterise the evolution of the system in terms of mean values and fluctuations. The fluctuations are further analysed using Lyapunov equation for the covariance matrix.
In \cref{III}, we derive an equation of motion for the effective dynamics of the two mechanical oscillators. This is achieved by performing an adiabatic elimination of the cavity modes.
In \cref{VII}, we investigate the correlations between the two mechanical modes through the evaluation of mutual information and quantum discord. Interestingly, we find that these asymmetric correlations (Discord) retain non-zero values even in the stationary state, indicating the potential for establishing persistent correlations.
In \cref{X}, we show how, in the cooling regime, the unidirectional coupling leads to the thermalization of the two mechanical modes at different temperatures, depending on the frequency mismatch of the mirrors. This reveals the so far unexplored setup to engineer a temperature gradient using the cascaded configuration.
\Cref{VIII} focuses on the analysis of the power spectra of the two mirrors and the output field mode, establishing the relationship between them.
In \cref{VI} we study the stability regions of the parameters space, exploring when the system can exhibit multistability. Finally in \cref{XII} we draw our conclusions.

\section{The system: two cascaded optomechanical cavities\label{sec2}}
\begin{figure}[b!]
    \centering
	\includegraphics[width=0.9\columnwidth]{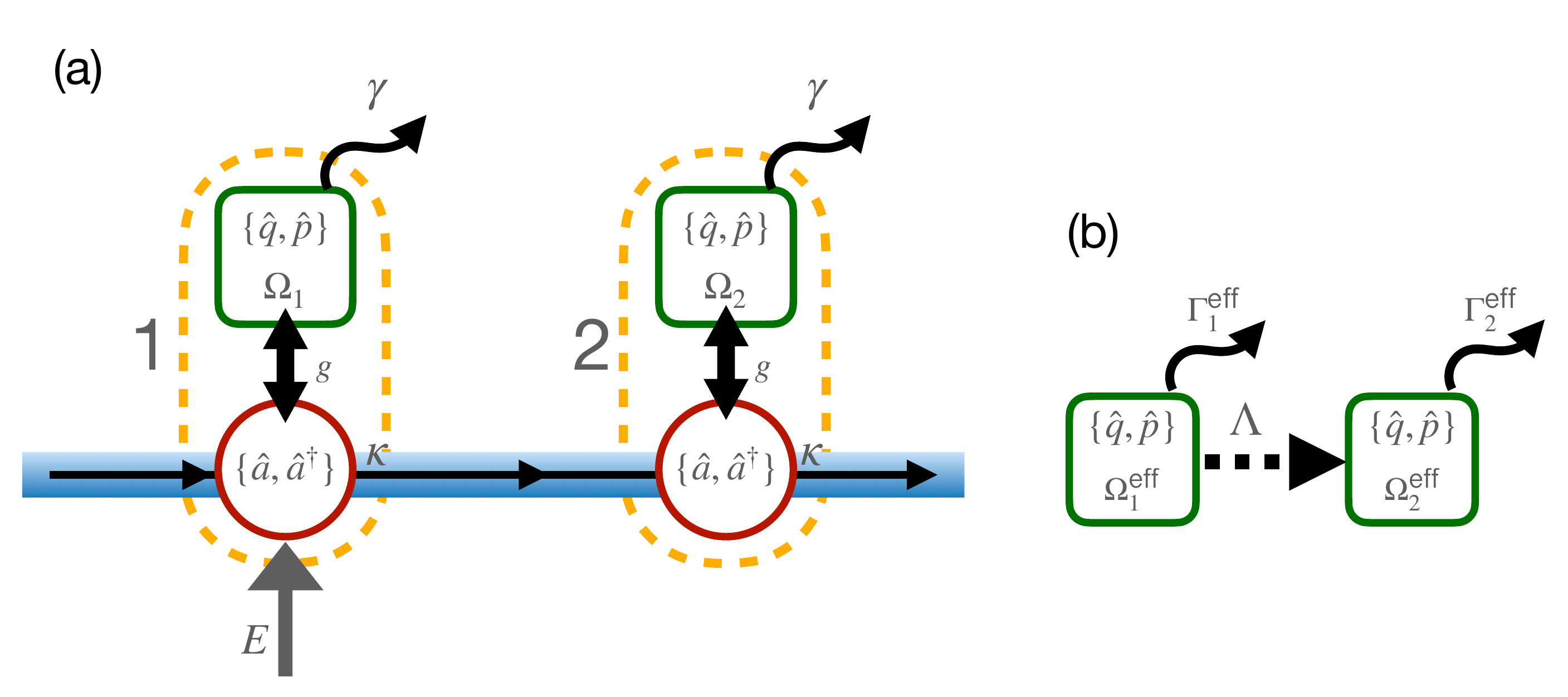}
	\caption{
 $(\mathbf{a})$ Sketch of the model involving two optomechanical systems: Each system consists of an optical mode and a mechanical mode, mutually interacting through the radiation pressure force induced by external laser power.
 We stress that these optomechanical systems are coupled to a unidirectional waveguide.
 In this context, the operators ${\hat{q},\hat{p}}$ represent the mechanical position and momentum, while ${\hat{a},\hat{a}^\dagger}$ are the creation and annihilation operators for the optical modes.
 The parameters $\gamma$ and $\kappa$ denote the mechanical and optical decay rates, respectively.
 Furthermore, $\Omega_{1,2}$ refer to the natural frequencies of the mechanical modes, and $g$ represents the coupling strength between the optical and mechanical modes due to radiation pressure. The amplitude of the coherent driving of the laser is denoted as $E$.
 ($\mathbf{b}$) Effective model describing the coupling between the mirrors, obtained by performing an adiabatic elimination of the optical modes. Within this framework, the effective frequencies of the mechanical modes are denoted as $\Omega^{\text{eff}}_{1,2}$, and $\Gamma^{\text{eff}}_{1,2}$ represents the effective decay rates. Moreover, $\Lambda$ (as defined in \cref{lambda}) represents the cascaded effective coupling between the two mechanical modes.}
	\label{disegno}
\end{figure}

Our system  consists of two optomechanical mirrors each coupled to the same unidirectional waveguide. Such mediated indirect interaction (see \cref{disegno}) leads to a cascade scenario in which the first system drives the following one without back action.
Each subsystem consists of a mechanical harmonic oscillator with mass $m$ and frequency $\Om_j$  coupled to a cavity field by means of its radiation pressure. As $\Om_j$ is much smaller than $c/2L$ ($c$ is the speed of light and $L$ stands for the cavity length) we can consider a single cavity mode \cite{law_interaction_1995,genes_chapter_2009} and write the following Hamiltonian for both subsystems:
\begin{equation} \label{H}
 \ham_S =\sum_{j=1}^2\ham_{j}=\sum_{j=1}^2\hbar\om_c \ajd\aj{+} \dfrac{\hbar\Omega_{j}}{2} (\qj^{2}{+}\pj^{2}) - \hbar g_j\ajd\aj \qj  + i\hbar E_{j}(\ajd e^{-i\om_L t}{-}\aj e^{i\om_L t})
\end{equation}
where $\aj$ is the cavity mode annihilation operator with optical frequency $\om_c$ of the $j$-th subsystem and $\qj$ ($\pj$)  stands for dimensionless position (momentum) operators of mechanical mode. The term proportional to $g_j{=}\om_c/L\sqrt{\hbar/(m\Om_{j})}$ describes the optomechanical coupling, while the last term is  the coherent input field with frequency $\om_L$. The quantities $E_j$ are related to the input powers $P_j$ by $E_j=\sqrt{2\kappa P_j/(\hbar\om_L)})$ where $\kappa$ is the cavity decay rate.
In a rotating frame at laser frequency $\om_L$, we define $\Delta=\om_c - \om_L$ and  \cref{H} becomes
\beq\label{H2}
\ham_S{=}\sum_j\hbar\left(\D - g_j \qj\right)\ajd\aj{+} \frac{ \hbar\Om_{j}}{2} (\qj^{2}{+}\pj^{2}) + i\hbar E_{j}(\ajd -\aj)
\eeq
In the following we focus on the scenario in which the dynamics of the second optomechanical system is driven only by the output field of the first cavity. Therefore, henceforth, we assume that the external laser  pumps the first cavity only (i.e. $E_2=0$).

Due to the inherently open nature of our system, we consider that each mechanical mode is coupled to its respective environment, which is assumed to be at a finite temperature \cite{giovannetti_phase-noise_2001}. Additionally, we account for photon leakage from the cavities. Specifically, we assume that the optical dissipation occurs through a unidirectional waveguide, resulting in a cascade-like coupling between the two optical modes mediated by the interaction with the guide \cite{karg_remote_2019}.

Following the standard input-output prescription of optical quantum Langevin equations \cite{gardiner_driving_1994}, we introduce the radiation vacuum input  noise operator  $\ain$ \cite{gardiner_quantum_nodate,gardiner_input_1985} and the Brownian noise operator $\hat \xi_j$ \cite{giovannetti_phase-noise_2001} with autocorrelation functions:
\begin{subequations}\label{autocorr}
\begin{align}
&\langle\ain(t)\aind(t')\rangle=\delta(t-t')\\
&\langle\{\hat\xi_j(t),\xi_j(t')\}\rangle=\ga_j (2 \bar{n}_j+1)\delta(t-t')
\end{align}
\end{subequations}
where $\gamma_j$ denotes the decay rate of the j-th mirror and $\bar{n}_j=1/(\exp(\hbar\Om_j/k_B T)+1)$ in which $k_B$ is the Boltzman constant and $T$ is the temperature of the mechanical modes' bath. Although the cavity and the resonator are at the same temperature, however, the cavity frequency is typically orders of magnitude larger than the mechanical frequency, therefore the average number of photons in the optical environment is negligible.
Based on the preceding analysis, we can derive the following quantum Langevin equations for the field operators $\aj$:
\begin{subequations}\label{opt_eq}
\begin{align}
	&\frac{d\ai}{dt}=-\frac{i}{\hbar}[\ai, H_{S}]-\frac{\kappa}{2}\ai-\sqrt{\kappa}\ain\\
	&\frac{d\aii}{dt}=-\frac{i}{\hbar}[\aii, H_{S}]-\frac{\kappa}{2}\aii-{\kappa}\ai-\sqrt{\kappa} \ain
\end{align}
\end{subequations}
and the mirror operators $\qj$, $\pj$
\begin{subequations}\label{mech_eq}
\begin{align} 
	&\frac{d\qj}{dt}=-\frac{i}{\hbar}[\qj, H_{S}]\\
	&\frac{d\pj}{dt}=-\frac{i}{\hbar}[\pj, H_{S}]-\ga_{j}\pj-\sqrt{\ga_{j}}\,\hat{\xi}_{j}
\end{align}
\end{subequations}

\emph{Mean field equations and fluctuations Dynamics} - 
The combined dynamics of the field-mirror system resulting from equations \cref{opt_eq} and \cref{mech_eq} is nonlinear. To investigate the quantum characteristics of optomechanical systems, a common method is to initially seek the mean field solution of the field and mechanical operators, and subsequently analyze the linearized dynamics of quantum fluctuations around these average values. Accordingly, we represent the operators as the sum of their average value (a $c$ number) and a small quantum fluctuation.
\beq\label{fluct}
\hat o=\langle \hat o \rangle+ [\hat o-\langle\hat o\rangle]=O+\fl{\hat o}
\eeq
This leads to the following set of non linear differential equations for the mean values 
\begin{subequations}\label{AVeq}
\begin{align}
    &\frac{dQ_j(t)}{dt}=\Om_j P_j(t)\\
    &\frac{dP_j(t)}{dt}=-\Om_jQ_j(t)-\ga_jP_j(t)+|G_j(t)|^2/g_j\\
    &\frac{dA_1(t)}{dt}=-\frac{\kappa}{2}A_1(t)-i\Delta_1(t) A_1(t)+E\\
    &\frac{dA_2(t)}{dt}=-\frac{\kappa}{2}A_2(t)-i\Delta_2(t)A_2(t)-\kappa A_1(t)
\end{align}
\end{subequations}
where we defined $\D_j(t){=}\D{-}g_jQ_j(t)$ and $G_j(t){=}g_j A_j(t)$. It is important to note the inherent asymmetry in comparison to the bidirectional case (cfr. \ref{bid}).
Solving \cref{AVeq} and using \cref{fluct} into \cref{opt_eq,mech_eq}, we can write the following linearized set of equations for the fluctuations, where we keep only terms $\mathcal{O}(\fl{\hat o})$:

\begin{subequations}\label{flu_eq}
\begin{align}
    &\frac{d\fl{\qj}}{dt}=\Om_j\fl{\pj}\\
	&\frac{d\fl{\pj}}{dt}=-\Om_j\fl{\qj}+(G_j^*\fl{\hat{a}}_j+G_j\fl{\hat{a}}_j^\dagger)-\ga\fl{\pj}-\xi_{j}\\
	&\frac{d\fl{\ai}}{dt}=iG_1\fl{\hat q_1}-i \D_1\fl{\ai} -\frac{\kappa}{2}\fl{\ai}-\sqrt{\kappa}\ain\\
	&\frac{d\fl{\aii}}{dt}=iG_1\fl{\hat q_2}-i \D_2\fl{\aii}-\frac{\kappa}{2}\fl{\aii}-\kappa\fl{\ai}-\sqrt{k} \ain
\end{align}
\end{subequations}

\emph{Covariance matrix and Lyapunov equation} - Given that the set of quantum Langevin \cref{flu_eq} is linear and  the quantum noise is Gaussian, we can fully characterize the quantum fluctuations dynamics in terms of the  
covariance matrix $\mathbf C$, whose elements are defined by $C_{ij}=1/2\langle \uop_i\uop_j{+}\uop_j\uop_i\rangle$ with $\vec{u}{=}\otimes_{j=1}^2\{\hat q_j,\hat p_j,\hat x_j,\hat y_j\}$ and the cavity field quadratures as	$\hat x{=}1/\sqrt{2}(\hat a{+}\hat a^\dagger)$, $\hat y{=}-i/\sqrt{2}(\hat a{-}\hat a^\dagger)$.
It follows that the $\mathbf C$ matrix obeys the following Lyapunov equation \cite{marquardt_quantum_2007,wilson-rae_theory_2007}:

\begin{equation}\label{Lyapunov}
\frac{d\mathbf C(t)}{dt}=\mathbf S(t)\mathbf C(t)+\mathbf C(t)\mathbf S(t){+}\mathbf N \;\; \text{with}\;\;
%\begin{align}%\label{SN}
\mathbf S{=}\begin{pmatrix}
	\mathbf S_1& \mathbf0\\
	\mathbf S_R& \mathbf S_2\end{pmatrix}
\text{and}\;\; 
\mathbf N{=}\begin{pmatrix}
		\mathbf N_1& \mathbf N_{12}\\
		\mathbf N_{12} &\mathbf N_2\end{pmatrix}
%\end{align}
\end{equation}\\
The block matrices entering in the drift ($\mathbf S$) and diffusion ($\mathbf N$) parts, reflect the unidirectionality of the model (as can be seen comparing them to the ones of the bidirectional case cfr. \ref{bid}):
%where the block matrices entering in $\mathbf S$ and $\mathbf N$ are given by:

\begin{equation}\label{block_matrices_S}
\mathbf S_j{=}\begin{pmatrix}
		0&\Om_{j} & 0 & 0 \\
		-\Om_{j}&-\ga & \text{Re}(G_j) & \text{Im}(G_j)\\
		-\text{Im}(G_j) & 0 & -\kappa/2 & \Delta_j\\
		\text{Re}(G_j) & 0 & -\Delta_j & -\kappa/2
		\end{pmatrix}
\qquad
\mathbf S_R{=}\begin{pmatrix}
	0&0 & 0 & 0 \\
	0&0 & 0 & 0 \\
	0&0 & -\kappa & 0 \\
	0&0 & 0 & -\kappa 
\end{pmatrix}
\end{equation}
and 
\beq\label{block_matrices_N}
\mathbf N_j{=}\begin{pmatrix}
	0&0 & 0 & 0 \\
	0&\ga(2 \bar n_j+1) & 0 & 0 \\
	0&0 & \kappa & 0 \\
	0&0 & 0 & \kappa \\
\end{pmatrix}
\;\;
\mathbf N_{12}{=}\begin{pmatrix}
	0&0 & 0 & 0 \\
	0&0 & 0 & 0 \\
	0&0 & \kappa & 0 \\
	0&0 & 0 & \kappa 
\end{pmatrix}
\eeq
%where $\Re G_j$ and $\Im G_j$ stand respectively for the real and the imaginary part of $G_j$.

\section{Effective mirrors dynamics}\label{III}

The optomechanical coupling, combined with the coupling of the cavity modes to the unidirectional waveguide, gives rise to an effective mediated interaction between the two mechanical modes. In the weak coupling regime ($G_j\lesssim \kappa$), we can explicitly derive an effective interaction by adiabatically eliminating the cavity field degrees of freedom associated with the cavities.
Indeed, if $G_j\lesssim \kappa$ we can focus on evolution of mechanical operators $\bar{b}_j$ and $\bar{b}_j^\dagger$, defined through $\fl{\qj}{=}(\bar{b}_je^{-i\Omega_j t}+\bar{b}_j^\dagger e^{i\Omega_j t})/\sqrt{2}$ and $\fl{\pj}{=}i(\bar{b}_j^\dagger e^{i\Omega_j t}-\bar{b}_j e^{-i\Omega_j t})/\sqrt{2}$.  
By considering these operators, we effectively eliminate the rapid timescale associated with the evolution of the cavity field \cite{Serafini}.
From \cref{flu_eq}, dropping counter-rotating terms, one obtains
\begin{align}\label{eqb}
    &\frac{d\bar{b}_j}{dt}=\dfrac{ie^{i\Om_j t}}{\sqrt{2}}(G_j^*\fl{\hat{a}}_j+G_j\fl{\hat{a}}_j^\dagger)-\dfrac{\ga}{2}\bar{b}_j-\dfrac{ie^{i\Om_j t}}{\sqrt{2}}\xi_{j}
\end{align}
The expressions for the cavity fields fluctuations can be found solving the respective equations in the frequency domain by using $\hat O(t) = 1/\sqrt{2\pi}\int_{-\infty}^{+\infty}\hat O(\om)e^{-i\om t}$. Therefore we rewrite the last two of the   \cref{flu_eq} as 
\begin{subequations}\label{flu_w_eq_1}
\begin{align}
    &\fl{\ai}(\om)=\chi_{a_1}(\om) \left(\ain(\om) \sqrt{\kappa}+i G_1 \fl{\hat q_1}(\om)\right)\\
    &\fl{\aii}(\om)=\chi_{a_2}(\om) \left(\ain(\om) \sqrt{\kappa}+i G_2 \fl{\hat q_2}(\om)-\kappa\fl{\ai}(\om)\right)
\end{align}
\end{subequations}
where we introduced the natural susceptibility of the optical modes $\chi_{a_j}$
\begin{align}\label{sus}
    &\chi_{a_j}(\om)=\frac{1}{\kappa/2-i \left(\omega -\Delta_j\right)}.
\end{align}

The optical input noise can be neglected as it is small compared to the mechanical thermal noise, given that the system evolves at room-temperature. So, back in time domain we obtain 

\begin{subequations}\label{flu_w_eq_2}
\begin{align}
    &\fl{\ai}(t)=\dfrac{i G_1}{\sqrt{2\pi}}\int d\om \chi_{a_1}(\om)\fl{\hat q_1}(\om)e^{-i\om t}\\
    &\fl{\aii}(t)= \dfrac{i G_2}{\sqrt{2\pi}} \int d\om\chi_{a_2}(\om)\left(\fl{\hat q_2}(\om)-\kappa\fl{\ai}(\om)\right)e^{-i\om t}
\end{align}
\end{subequations}
Thanks to the properties of convolutions in Fourier transforms,  assuming that $\bar{b}_j$ and $\bar{b}^\dagger_j$ vary slowly in time, \cref{flu_w_eq_2} become
\begin{subequations}
\begin{align}
    \fl{\ai}(t)=&\dfrac{i G_1}{\sqrt{2}}\left(\bar{b}_1 \chi_{a_1}(\Om_1)e^{-i\Om_1 t}{+}\bar{b}^\dagger_1 \chi_{a_1}^*(\minus{\Om_1})e^{i\Om_1 t}\right)\\
    \fl{\aii}(t)=& \dfrac{i G_2}{\sqrt{2}}\left(\bar{b}_2 \chi_{a_2}(\Om_2)e^{-i\Om_2 t}{+}\bar{b}^\dagger_2 \chi_{a_2}^*(\minus{\Om_2})e^{i\Om_2 t}\right)-\nonumber\\
    &\dfrac{i \kappa G_1}{\sqrt{2}}\left(\bar{b}_1 \chi_{a_1}(\Om_1)\chi_{a_2}(\Om_1)e^{-i\Om_1 t}{+}\bar{b}^\dagger_1 \chi_{a_1}^*(\minus{\Om_1})\chi_{a_2}^*(\minus{\Om_1})e^{i\Om_1 t}\right)
\end{align}
\end{subequations}

By substituting these results into \cref{eqb} and neglecting non-rotating terms, we derive the following coupled equations of motion for the mirrors operators only:
\begin{subequations}
\begin{align}
        &\frac{d\bar{b}_1}{dt}=-i\Om^\text{eff}_1\,\bar{b}_1-\left(\Gamma^{\text{eff}}_{1}+\dfrac{\ga}{2}\right)\bar{b}_1-\dfrac{ie^{i\om_1 t}}{\sqrt{2}}\xi_{1}\\
        &\frac{d\bar{b}_2}{dt}=-i\Om^{\text{eff}}_{2}\,\bar{b}_2-\left(\Gamma^\text{eff}_{2}+\dfrac{\ga}{2}\right)\bar{b}_2-\frac{\kappa}{2}\Lambda(\Om_1)\bar{b}_1-\dfrac{i e^{i\om_2 t}}{\sqrt{2}}\xi_{2}
\end{align}
\end{subequations}

in which $\Gamma^{\text{eff}}_{j}$  and $\Om^{\text{eff}}_{j}$ are the effective (induced by the adiabatically eliminated optical fields) decay rate and  mechanical frequencies, defined respectively as the real and imaginary part of $ |G_j|^2\left(\chi_{a_j}(\Om_j){-}\chi_{a_j}^*(\minus{\Om_j})\right)/2$, and $\Lambda(\Om_1)$ is the effective cascaded coupling between the two mirrors (see \cref{disegno}), defined by
\beq\label{lambda}
\Lambda(\om){=}G_2G^*_1\chi_{a_1}^*
    (\minus{\om})\chi_{a_2}^*(\minus{\om}){-}G^*_2G_1\chi_{a_1}(\om)\chi_{a_2}(\om).
\eeq
These equations can be recast in a covariance matrix equation form (cfr. \cref{Lyapunov}) with $\vec{u}{=}\otimes_{j=1}^2\{\bar{b}_j,\bar{b}^\dagger_j\}$. In this case, the blocks entering in the drift $\mathbf S$ and diffusion $\mathbf N$ matrices turn out to be
\beq
\mathbf S_j{=}\begin{pmatrix}
		-i\Om^\text{eff}_j-(\Gamma^\text{eff}_{j}+\dfrac{\ga}{2})&0\\
		0&i\Om^\text{eff}_j-(\Gamma^\text{eff}_{j}+\dfrac{\ga}{2})\\
		\end{pmatrix} \qquad
\mathbf S_R{=}\begin{pmatrix}
	-\Lambda(\Om_1)&0 \\
	0&-\Lambda(\Om_1)^*\\
\end{pmatrix}
\nonumber
\eeq
and 
\begin{align}
\mathbf N_j{=}\begin{pmatrix}
	\ga(2 \bar n_j+1)&0  \\
	0&\ga(2 \bar n_j+1) \\
\end{pmatrix}
\qquad
\mathbf N_{12}{=}\begin{pmatrix}
	0&0\\
	0&0\\ 
\end{pmatrix}
\nonumber
\end{align}

Notice that the noise matrix $\boldsymbol{N}$ is diagonal due to the fact that we have dropped counter-rotating terms.
So, as said in the beginning, we have found that, looking on a timescale larger than the one describing the cavities modes evolution, mirror's interaction mediated by the cavities can be described as an effective coupling.  

\section{Mutual Information and Quantum Discord}\label{VII}
Once \cref{Lyapunov} is solved, we can analyse and conveniently characterise the  mirrors correlations - both in the transient and in stationary regimes - by means of the mutual information, which can be evaluated from the covariance matrix, as shown in \cite{olivares_quantum_2012}, in terms of its symplectic invariants and  symplectic eigenvalues (see \ref{AppA}). 

\begin{figure}[h!]
    \centering
    \includegraphics[width=1\columnwidth]{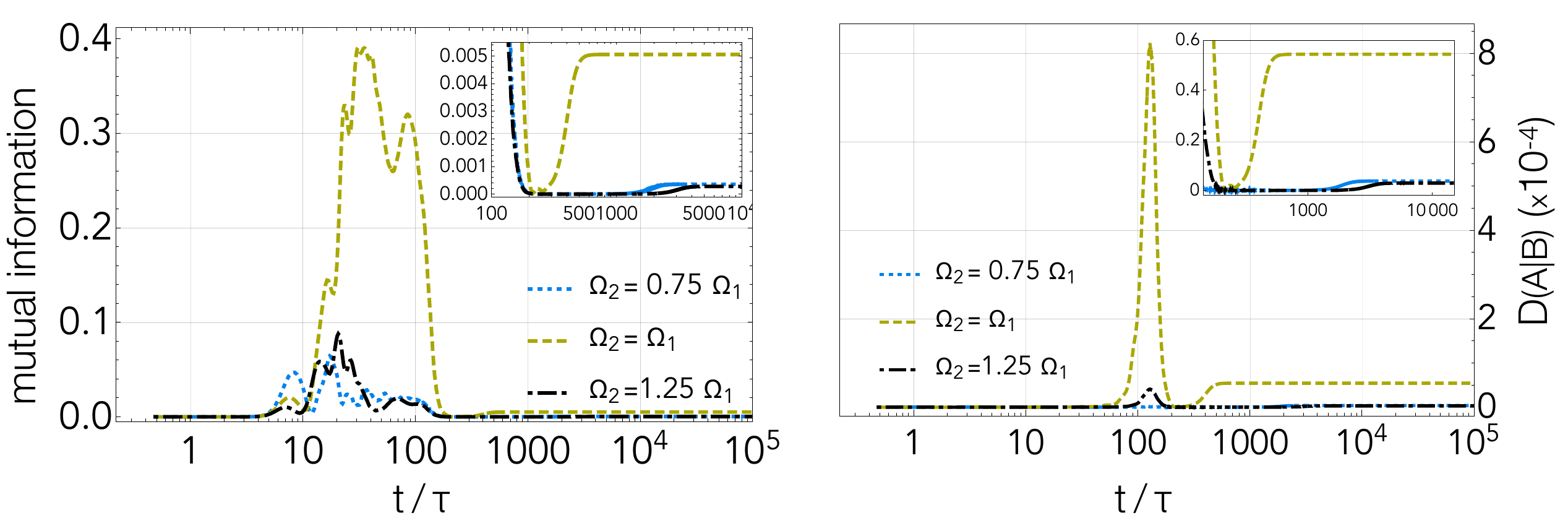}

	\caption{(Left) Mutual information between the two mirrors as a function of time in the cooling regime ($\Delta=\Om_1$). The dotted line refers to the case $\Om_2=0.75 \Om_1$, the dashed line refers to the case $\Om_2=\Om_1$ and the dot-dashed one refers to the case $\Om_2=1.25\Om_1$. The box in the right corner zooms on the long time region where is shown the establishment of steady state correlations (Right) Quantum discord $D(A|B)$ in the same cases considered for the mutual information. The box in the right corner shows a zoom of the discord values on the long time region. For these plots the following parameters' values have been considered:
$m {= }150 $ ng, $\Om_1 /(2\pi ){=}1$ MHz, $\ga/(2\pi){=}1$ Hz (\textcolor{red}{$\ga_j$}), $T {=}300$ K, $L{ =}25$mm, $\kappa{=}1.34$ MHz, $\lambda{=}1064$ nm, and $P_1{ =}2$ mW and time is expressed in units of $\tau=2\pi/\Om_1$.  These values are consistent with the state of the art experiments and unless otherwise specified these are the ones considered in all the shown results}
	\label{fig14}
\end{figure}

As shown in \cref{fig14},  the time evolution of the mutual information is characterised by distinct phases. Initially, the two mirrors exhibit no correlation, indicating an uncorrelated state. Following this, a short transient of correlation between the mirrors occurs. Eventually, they return to an uncorrelated state again, only to reach a state of steady-state correlations. Comparing this result with the temperature behaviour reported in \cref{fig13} can be observed that the rebirth of correlations occurs subsequent to thermalization of the second mirror.
The amount of quantumness and, most importantly, the asymmetrical nature (due to the unidirectional coupling) of such correlations can be characterized in terms of quantum Discord \cite{Zurek2001}.

This different type of quantum correlations can be nonzero even in the case of separable states which implies that  some bipartite quantum states can show correlations that are incompatible with classical physics. For our system we can adopt the Gaussian  quantum Discord \cite{Adesso2010,Giorda2010}. Quantum Gaussian Discord is defined as the difference between mutual information and classical correlations. Classical correlations are defined as the maximum amount of information that one can gain on one subsystem by locally measuring the other subsystem \cite{olivares_quantum_2012} and so, by this definition, quantum Discord is not symmetric with respect to the interchange of the two subsystems. In particular in this case, where the coupling is unidirectional, the Gaussian discord is maximally asymmetrical due to the fact that measuring the first subsystem one cannot recover any information about the second one.

The quantum Discord $D(A|B)$ plotted in \cref{fig14} refers only to the one relative to the second subsystem conditioned to the first, indeed performing a measure on the second mirror one can recover some information on the first one, but the converse is not true. In fact $D(B|A)$ is identically zero at all times. That is expected due to the unidirectionality of the coupling. Furthermore also for the quantum discord 
there is a non zero value also in the stationary state.

\section{Finite temperature gradient}\label{X}
Here we will show that in the cooling regime, due to the effective unidirectional coupling, a temperature gradient is established between the two mechanical modes. In fact in the cooling regime, i.e. $\Delta{=}\Omega_1$, the optical field generates extra damping on the mechanical mode. Such optical damping, caused by radiation pressure, depends on both the  position $Q$ and  the speed with which the mirror changes its position.
At $t=0$ the phonons associated to the mechanical oscillator motion are in a thermal equilibrium state. Then, the interaction between the photons and the phonons, as described by the last term in \cref{H2}, leads to a change of the phonon number which fluctuates due to the coupling to its environment, consisting of a hot phonon bath at temperature $T$. The goal of optomechanical (sideband) cooling is to reduce the amount of such fluctuations thereby cooling it down.

The mean energy of the mirrors is evaluated 
\begin{align}
    U_j(t)=\frac{\hbar \Om_j}{2}\left(\langle\fl{q}_j^2(t)\rangle+\langle\fl{p}_j(t)^2\rangle\right)=\hbar \Om_j\left(n^{\text{eff}}_j(t)-\frac{1}{2}\right)
\end{align}
with $n^{\text{eff}}_j(t)$ obtained by the solution \cref{Lyapunov} of covariance matrix  as $1/2(C_{11}+C_{22}-1)$ for the first mechanical mode and $1/2(C_{55}+C_{66}-1)$ for the second one. 
\begin{figure}[h!]\centering
\includegraphics[width=0.8\columnwidth]{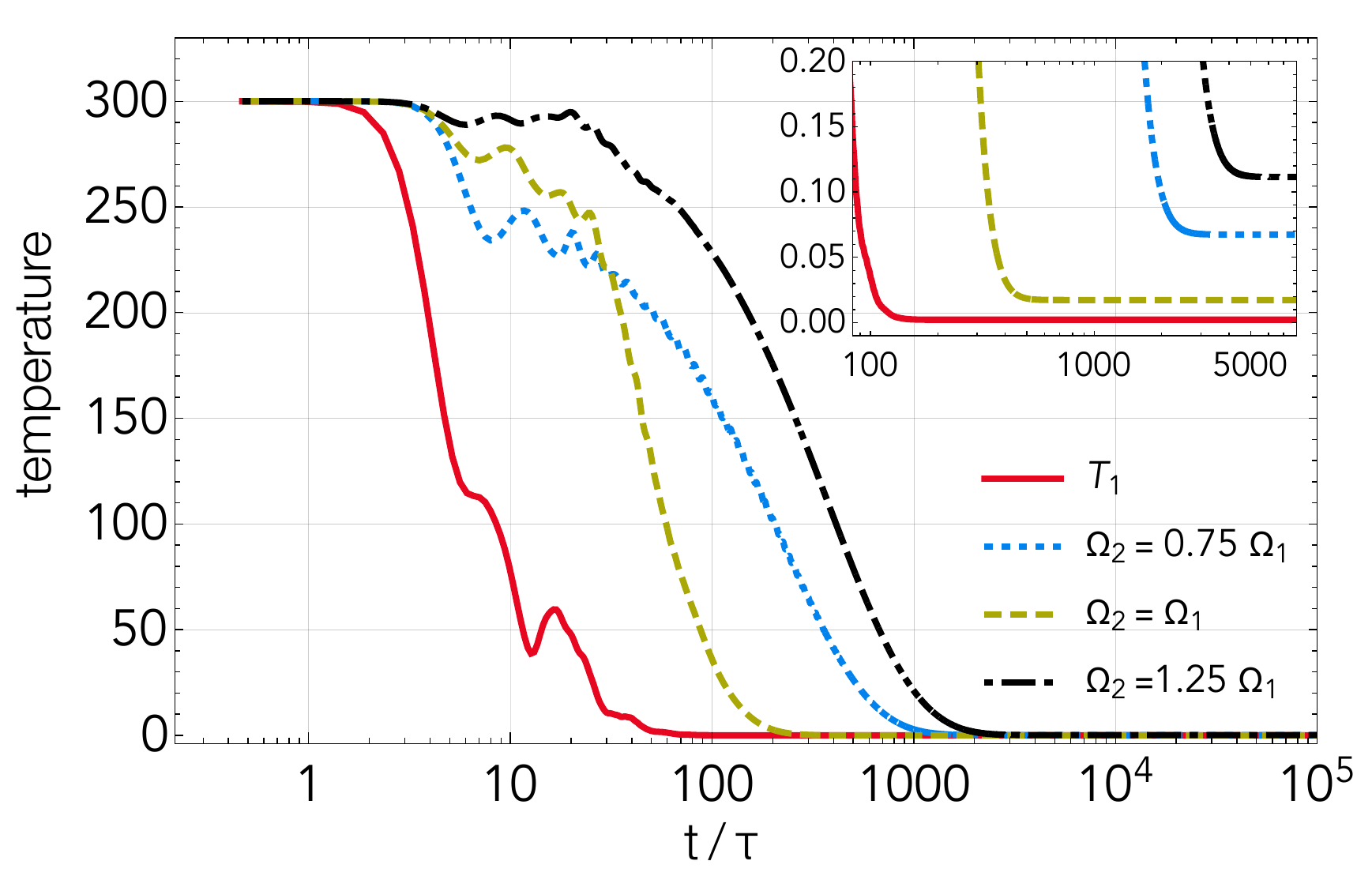}
	\caption{Temperatures of mechanical modes, as defined in \cref{Temp}, as a function of time. The solid line refers to the temperature of the first mirror, the dotted one refers to the temperature of the second mirror in the case $\Om_2=0.75 \Om_1$, the dashed line represents the the temperature of the second mirror in the case $\Om_2=\Om_1$, while the dot-dashed one represents the temperature of the second mirror in the case $\Om_2=1.25 \Om_1$}	
	\label{fig13}
\end{figure}

The effective temperature of the movable mirrors are then given by
\begin{align}\label{Temp} T^{\text{eff}}_j(t)=\frac{\hbar\Om_j}{k_B\ln{(1+1/n^{\text{eff}}_j(t)})}
\end{align}
When the two mirrors have the same frequency, i.e. $\Omega_2=\Omega_1$ (\cref{fig13}), the steady state is characterised by a higher temperature of the second mirror with respect to the first one. This is a consequence of the unidirectionality of the coupling. Indeed, as shown in appendix, such temperature gradient is absent in the bidirectional case.

\begin{figure}[h!]
    \centering
	\includegraphics[width=1\columnwidth]{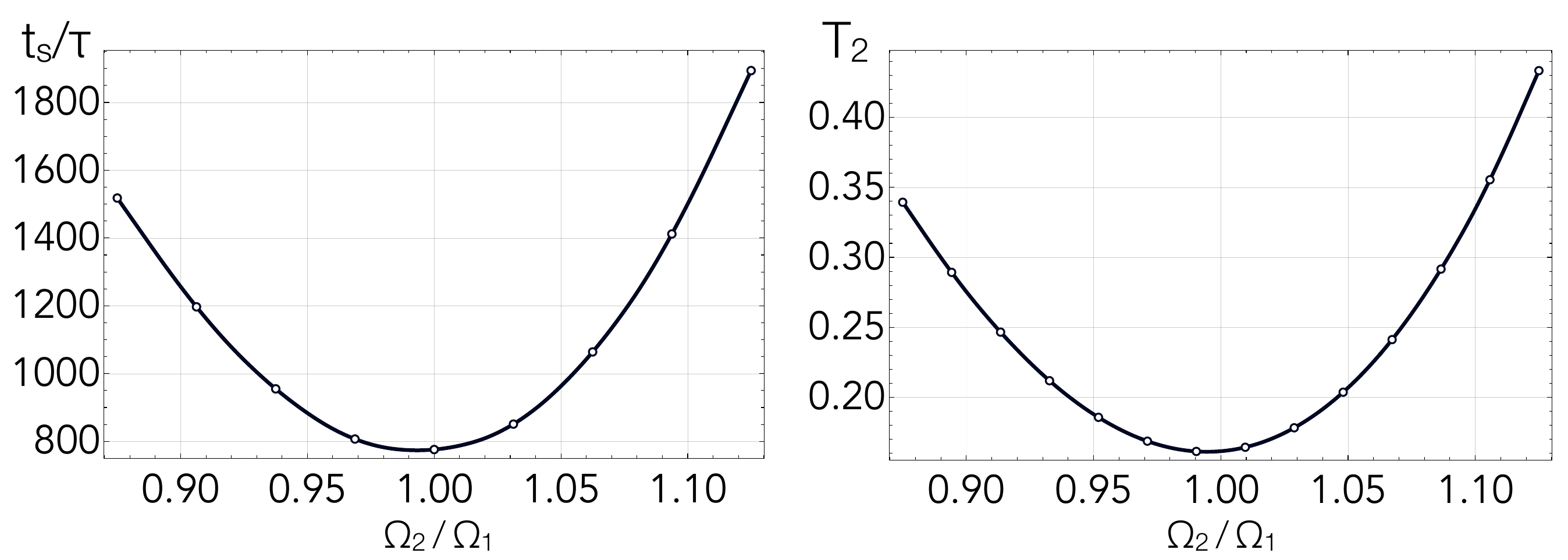}
	\caption{(Left) Time at which the second mechanical mode reaches the stationary state as a function of its frequency (Right)
	Temperature of the second mechanical mode in the stationary regime as a function of its frequency }
	\label{fig15}
\end{figure}

To further investigate the properties of this temperature gradient, the temperatures of the second mirror were evaluated varying its frequency. Due to the mismatch between the optical detuning and the frequency of the mechanical mode in the second optomechanical system, which corresponds to a variation in the cooling efficiency, the time at which the second mirror reaches a steady temperature value, $t_s$, increases as shown in \cref{fig15} as long as the frequency is moving apart from the resonant case $\Om_2=\Om_1$ 

\begin{figure}[h!]
    \centering
\includegraphics[width=1\columnwidth]{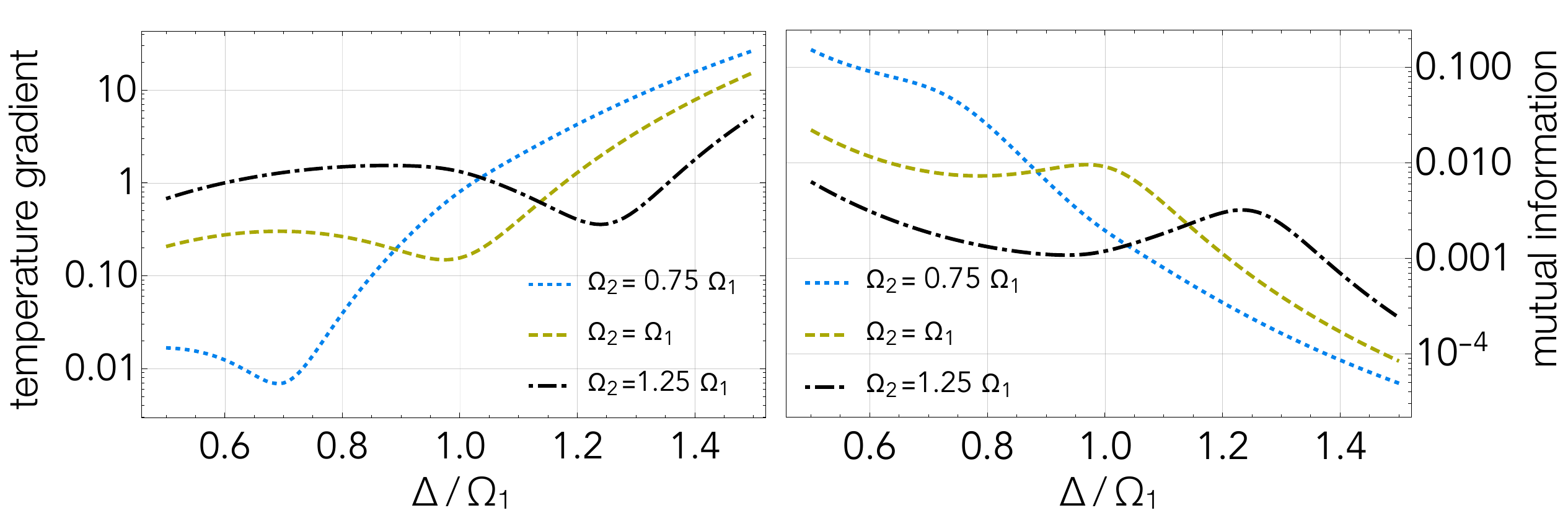}
	\caption{(Left) Temperature gradients of mechanical modes $T^\text{eff}_2-T^\text{eff}_1$, where $T^\text{eff}_j$ is defined as in  \cref{Temp}, in the stationary limit for different values of $\Om_2$. In particular, the curves refer to the cases $\Om_2=0.75\Om_1$, $\Om_2=\Om_1$ and $\Om_2=1.25\Om_1$. (Right) Mutual information between the two mechanical modes in the stationary limit for the same values of $\Om_2$}
	\label{temps}
\end{figure}

It can be seen in  \cref{temps} that for different values of $\Omega_2$, varying the detuning between the pump and the first cavity, one can always tune it in such a way that it creates a temperature gradient between the mirrors. The stationary correlations between the two mirrors, evaluated as the mutual information in the stationary regime, shows a peak in correspondence to the minima of the second mirror temperatures.

\section{The steady state and the power spectra}\label{VIII}
One of the experimentally accessible quantity for the optomechanical systems is the power spectrum of the cavity output field  which allows to reconstruct the spectrum (and so the dynamics and the temperature) of the mechanical mirror \cite{paternostro_reconstructing_2006}. We now show how in this cascaded configuration, the spectrum in output from the last cavity contains informations about the two mirrors and allows to reconstruct their dynamics. 
In order to evaluate the spectra of the cavity output and the two individual mirrors, it's necessary to evaluate the stationary state of the system. As shown in the previous sections, the linearized equations for the fluctuations \cref{flu_eq} can be solved in the frequency domain.
The correlation functions \cref{autocorr} become
\begin{subequations}\label{w_autocorr}
\begin{align}
&\langle\ain(\om)\aind(\om')\rangle=\delta(\om+\om')\\
&\langle\{\hat\xi_j(\om),\xi_j(\om')\}\rangle=2\ga \coth\left(\frac{\h\Om_j}{2k_BT}\right)\delta(\om+\om'), 
\end{align}
\end{subequations}
while the equations for the fluctuations of cavity field modes are \cref{flu_w_eq_1} and the equations for the positions of the mirrors become
\begin{align}\label{flu_w_eq}
    \fl{\qj}(\om)=\chi_j(\om) \left(G_j \fl{\ajd}(\om)+G^*_j\fl{\aj}(\om) +\xi_j(\om)\right)
\end{align}
where we have introduced the natural susceptibilities of the mechanical modes
%and the optical modes ($\chi_{a_j}$).
\begin{align}
     &\chi_j(\om)=\frac{\Om_j}{\Om_j^2-\om^2-i\om\ga}
%    \\
%    &\chi_{a_j}(\om)=\frac{1}{\kappa/2-i \left(\omega -\Delta_j\right)}
\end{align}

The mirror's position fluctuations can be expressed in terms of effective susceptibilities and noise operators:
\begin{subequations}\label{w_flu}
\begin{align}
   \fl{\hat q_1}(\om)= &\chi^{\text{eff}}_1(\om)
   %\; \; 
   \xi_1(\om)\\
   \fl{\hat q_2}(\om)= &\chi^{\text{eff}}_2(\om)
   %\left(\xi_2(\om)-i \kappa \chi^{\text{eff}}_1(\om)\xi_1(\om)\eta(\om)\right)
   \xi^{\text{eff}}_2(\om)
\end{align}
\end{subequations}
with
\begin{subequations}\label{chi_eff}
\begin{align}
    &\chi_j^{\text{eff}}(\omega)=\frac{\chi_j(\om)}{1- i|G_j|^2\chi_j(\om) (\chi_{a_j}(\om)-\chi_{a_j}^*(\minus{\om}))}\\
    &\xi^{\text{eff}}_2(\om)= \xi_2(\om)-i \kappa \chi^{\text{eff}}_1(\om)\xi_1(\om)\Lambda(\om)
\end{align}
\end{subequations}
Note that the effective susceptibility of the mechanical oscillators $\chi_j^{\text{eff}}(\omega)$ are modified by the radiation pressure \cite{genes_ground-state_2008}. Furthermore, in the second of   \cref{w_flu}, the effective noise seen by the second mirror, modified by the presence of the first, is made explicit.
It is now clear how the position fluctuations of the first ($\fl{\hat q_1}$) of the two mechanical modes depends only on its local thermal bath, while the second one ($\fl{\hat q_2}$)  depends also on the thermal bath of the first via the optical field.

In the same way, for the cavity field fluctuation we have
\begin{subequations}\label{w_cav_  }
\begin{align}
    \fl{\ai}(\om)&=i G_1  \chi_{a_1}(\om) \chi^{\text{eff}}_1(\om)\xi_1(\om)\\
    \fl{\aii}(\om)&=i G_2  \chi_{a_2}(\om) \chi^{\text{eff}}_2(\om)\xi^{\text{eff}}_2(\om)- i G_1 \kappa  \chi_{a_1}(\om) \chi_{a_2}(\om) \chi^{\text{eff}}_1(\om)\xi_1(\om)
\end{align}
\end{subequations}

\subsection*{Power Spectra}
From \cref{w_flu,w_cav_  }, thanks to \cref{w_autocorr}, it is possible to evaluate the position spectrum of the two mirrors defined by
\begin{align}
    S^q_j(\om)=\frac{1}{\sqrt{2\pi}}\int_{-\infty}^{+\infty}d\Om e^{-i(\om+\Omega)t}\av{\fl\qj(\om)\fl\qj(\Omega)}
\end{align}
obtaining
\begin{subequations}\label{Sq}
\begin{align}
    S^q_1(\om){=}&\ga(2\bar n_1{+}1)|\chi_1^{\text{eff}}(\om)|^2\\
    %S^q_2(\om)=&\ga(2\bar n_2+1)|\chi_2^{\text{eff}}(\om)|^2\nonumber\\
    %&-\ga(2\bar n_1+1)\kappa^2|\chi_2^{\text{eff}}(\om)|^2|\chi_1^{\text{eff}}(\om)|^2|\eta(\om)|^2\nonumber
    S^q_2(\om){=}&\ga(2\bar n_2{+}1)|\chi_2^{\text{eff}}(\om)|^2{-}\kappa^2 S^q_1(\om)|\chi_2^{\text{eff}}(\om)|^2|\Lambda(\om)|^2
\end{align}
\end{subequations}
from which one can obtain the variances $\av{\fl{\qj}^2}$ trough
\begin{align}
    \av{\fl{\qj}^2}=\int_{-\infty}^{+\infty}\frac{d\om}{2\pi}S^q_j(\om)
\end{align}

We are also interested to the output power spectral density that would be detected in an homodyne detection of the output fluctuations $\fl{x}^{\text{out}}{=}1/\sqrt{2}(\fl{\hat a}^{\text{out}}+\fl{\hat a}^{\dagger\text{out}})$ where 
\begin{align}
    \fl{\hat a}^{\text{out}}(\om)=\ain(\om)-\sqrt{\kappa}\fl\ai(\om)-\sqrt{\kappa}\fl\aii(\om)
\end{align}

The spectrum of such fluctuations can be obtained as
\begin{align}
    P^{\text{out}}(\om)=\frac{1}{\sqrt{2\pi}}\int_{-\infty}^{+\infty} d\Om e^{-i(\om+\Omega)t}\braket{x^{\text{out}}(\om)x^{\text{out}}(\Omega)}
\end{align}

\begin{figure}[h!]
    \centering
	\includegraphics[width=1\columnwidth]{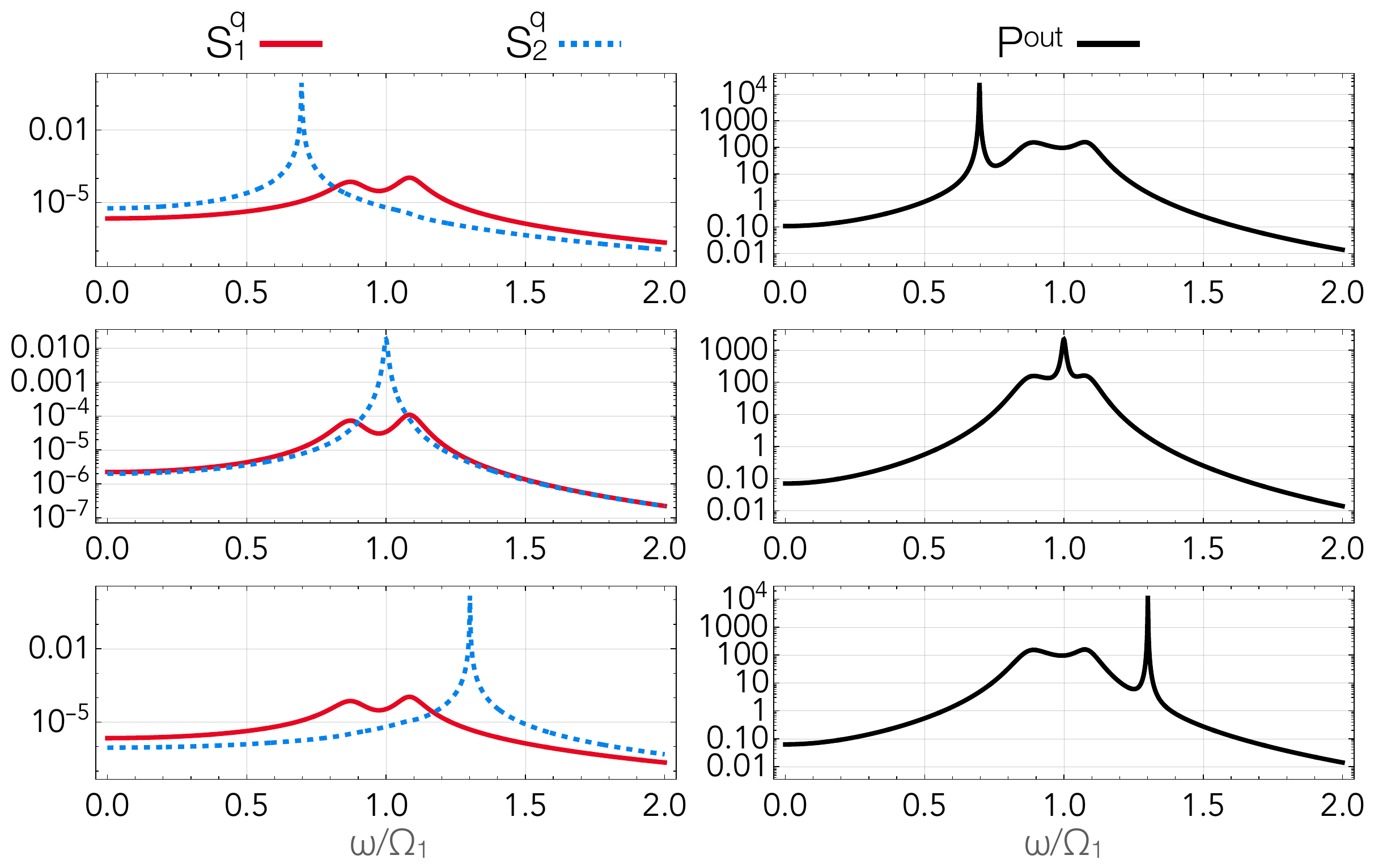}
	\caption{(Left) \emph{Mirror's Spectra} - In this figure we plot the power spectrum \cref{Sq} of the first mechanical mode and of the second one for three different frequencies, from top to bottom, $\Om_2=\Om_1/2$, $\Om_2=\Om_1$ and $\Om_2=3\Om_1/2$. (Right) \emph{Output Spectra} - Spectrum of the output field from the second cavity. In this case $P=10^{-2}mW$}	\label{spectra}
\end{figure}

Using \cref{w_autocorr} in the frequency domain one finds 
\begin{align}
	    %P^{\text{out}}(\om)=& 
	    %\kappa|\alpha(\omega)|^2 S^q_1(\omega){+}
	    %\kappa|\beta(\omega)|^2 S^q_2(\omega) {-}\nonumber\\
	   %& 2\kappa^2 \Im\left[\alpha(\omega)\beta(\omega)\eta(\omega)\chi^{\text{eff}}_2(\omega)\right] S^q_1(\omega)
        P^{\text{out}}(\om)\sim \sum_{j=1}^2
        K_j(\om) S^q_j(\om)
        \label{out}
\end{align}

with $K_j(\om)=\kappa\vert{G_j\chi_{a_j}(\om){-}G_j^*\chi_{a_j}^*(\minus{\om})}\vert^2/2$. From  \cref{out}, it follows, as shown  in \cref{spectra}, that the output field from the second cavity contains information on the power spectra of both mechanical modes as it simply proportional to the sum of the two mechanical power spectra. A similar result was obtained, for a single optomechanical system, in \cite{paternostro_reconstructing_2006}. Consequently, the peaks observed in the cavity's output spectrum reflect those of the individual mirrors' spectra. Hence, one can infer the power spectrum of each individual mirror by fitting the peaks of the spectrum, given that the power spectrum of an individual optomechanical mirror is well approximated by a Lorentzian curve \cite{bowen_quantum_2020}.

The appearance of two peaks in the power spectra of the first mirror, as depicted in \cref{spectra}, is contingent on the selected value for the first cavity's pump power. Specifically, these dual peaks manifest at a particular power pump value and progressively move farther apart as the power value increases.

\section{Self-induced oscillations and multistability}\label{VI}
Up until this point, our analysis has primarily focused on stable states by employing specific parameter values. However, we will now delve into the nonlinear regime for the average values and demonstrate how to identify stable states by tuning the system's parameters.
As the optomechanical coupling becomes stronger and damping becomes weaker, the system's nonlinearities become significant and cannot be neglected any longer. In this regime, the system exhibits instabilities, leading the mirror to enter a state of what is known as "self-sustained oscillations". In the following, we will explore these nonlinear dynamics and outline the conditions required to achieve stable states amidst the presence of these oscillations\cite{ludwig_optomechanical_2008,marquardt_dynamical_2006}. In this regime the mean position of the mirrors can be written as $Q_j(t)=\bar{Q}_j{+}\alpha_j cos(\Omega_j t)$. Putting this into \cref{AVeq}, the exact solutions for the cavity modes amplitude $A_j$, in the long time limit, can be written as 

\begin{subequations}
\begin{align}
    A_1(t)=&\exp\left[ig_1 \alpha_1\frac{\sin(\Omega_{1} t)}{\Omega_{1}}\right]
    %e^{i\varphi_1(t)}
    \sum_n A_{1}^ne^{i\Omega_1 n t}\\
    A_2(t)=&\exp\left[ig_2 \alpha_2\frac{\sin(\Omega_{2} t)}{\Omega_{2}}\right]
    %e^{i\varphi_2(t)}
    \sum_{n,m,l} A_2^{nml}e^{i(\Omega_{1}(n+l)+\Omega_{2}m) t}
\end{align}
\end{subequations}

with 

\begin{subequations}
\begin{align}
&A_1^n=J_n\left(\frac{-g_1 \alpha_1}{\Omega_{1}}\right)\chi_{a_1}(-\Om_1 n)E_1\\
&A_2^{nml}=-\kappa\; J_n\left(-\frac{g_1 \alpha_1}{\Omega_{1}}\right)\;J_m\left(-\frac{g_2 \alpha_2}{\Omega_{2}}\right)J_l\left(\frac{g_1 \alpha_1}{\Omega_{1}}\right)\chi_{a_2}(-\Om_1(n+l)-\Om_2 m)\; E_1
\end{align}
\end{subequations}
where $J_n(x)$ is the Bessel function of first kind and $\chi_{a_j}$ are the susceptibilities defined in \cref{sus}.
The stable states of the system are those for which the total time-averaged force vanishes and the power due to the radiation pressure $P_{rad_j}=G\langle|A_j|^2 \dot{Q}_j\rangle$ equals the power dissipated $P_{fric}=\gamma \langle \dot{Q}^2_j\rangle$. By plotting the ratio $P_{rad}/P_{fric}$ for the two subsystems as a function of $A_j$ and detuning $\Delta$, we obtain diagrams that illustrate the parameter regions corresponding to stable states. These diagrams provide valuable insights into the values of $A_j$ and $\Delta$ where the system exhibits stability.
\begin{figure}[h!]
    \centering
\includegraphics[width=0.9\columnwidth]{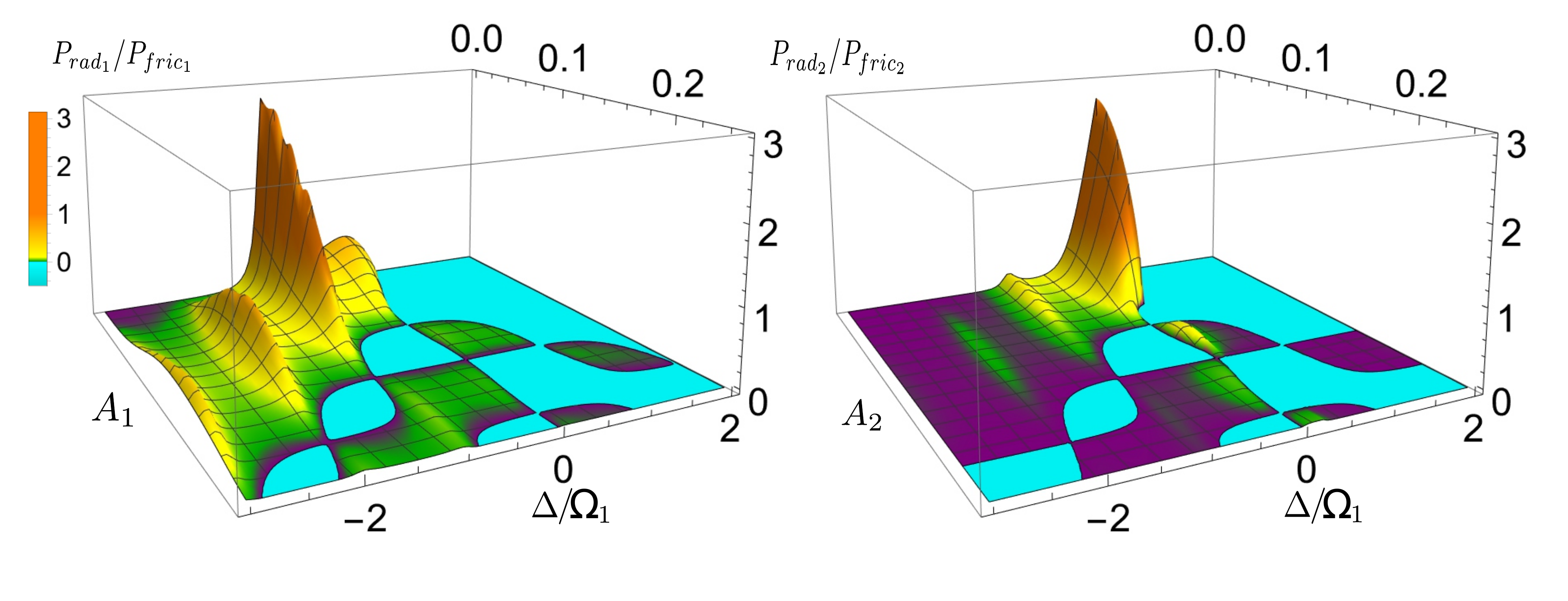}
	\caption{(Left) Stability plot for the first mechanical oscillator. It shows the ratio between the power due to radiation pressure and the power dissipated as a function of the amplitude of oscillation and the detuning between the pump and the cavity\cite{marquardt_dynamical_2006}. (Right) Stability Graph for the second mechanical oscillator, it shows the ratio between the power due to radiation pressure and the power dissipated as a function of the amplitude of oscillation and the detuning between the pump and the cavity. In both the plots, the stable states are those for which the power due to the radiation pressure $P_{rad_j}=G\langle|A_j|^2 \dot{Q}_j\rangle$ equals the power dissipated $P_{fric}=\gamma \langle \dot{Q}^2_j\rangle$}
	\label{stab}
\end{figure}

\begin{figure}[h!]
	\includegraphics[width=1.\columnwidth]{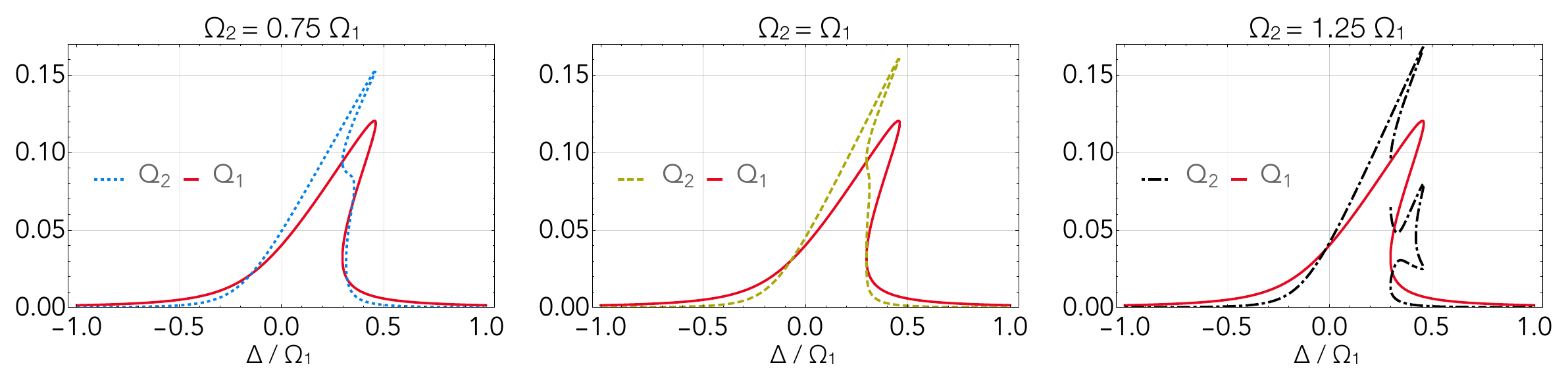}
	\caption{Steady-state position of the cantilevers as a function of detuning for $\Om_2{=}3/4\Om_1$(left), $\Om_2{=}\Om_1$(center) and  $\Om_2{=}5/4\Om_1$(right). As known in literature \cite{bowen_quantum_2020,Sarma:16}, the positions of mirrors show multistability. In a single optomechanical system the mechanical position have three stable solutions, but due to the unidirectional coupling the second one shows three or more stable solutions depending on the detuning between the two mirror's frequencies  }
	\label{fig3a}
\end{figure}

\emph{Multistability} - A characteristic feature of optomechanical systems is that, in the regime in which $A_j=0$, they exhibit multistability. A given intensity of the light pumped in the cavity can lead to different steady states of both cavity photon number and mechanical position \cite{bowen_quantum_2020,Sarma:16}. From \cref{AVeq}, taking the stationary limit, we can find the equations for the average number of photons in the two cavities $N_j$ i.e.
\begin{subequations}\label{AVphot}
\begin{align}
   &\dfrac{g^4_1}{\Omega^2_{1}}N^3_1-\dfrac{2\Delta g^2_1}{\Omega_{1}}N^2_1+\left(\Delta^2+\dfrac{\kappa^2}{2}\right) N_1-E^2_1=0\\
    &\dfrac{g^4_2}{\Omega^2_{2}}N^3_2-\dfrac{2\Delta g^2_2}{\Omega_{2}}N^2_2+\left(\Delta^2+\dfrac{\kappa^2}{2}\right) N_2-\kappa^2N_1=0\\
\end{align}
\end{subequations}
and once found these, we can find the average cantilever positions as 
\begin{equation}
    Q_j=\dfrac{g_j}{\Omega_j}N_j
\end{equation}
We note that the first of \cref{AVphot} has three roots, but, as shown also in \cite{Sarma:16}, only two of these solutions are stable solutions, specifically the lower and the higher ones while the middle one is unstable and can't be observed experimentally. Regarding the equation for the second cavity a richer behaviour is obtained, as shown in \cref{fig3a}.

\section{Conclusion}\label{XII}
In summary, this study has provided a comprehensive characterization of the dynamics of two optomechanical systems indirectly coupled via a chiral waveguide in a cascaded configuration. In the weak coupling regime, we employed an adiabatic elimination technique to derive effective equations governing the mirror dynamics.

By examining the evolution of correlations between the two mechanical modes, we quantified their mutual information and quantum discord over time. Remarkably, our results demonstrate that these correlations persist even in the stationary state, indicating a non-zero value of steady correlations.
Furthermore, we investigated the steady-state temperatures of both mirrors for different values of $\Omega_2$ and various $\Delta$ parameters. Our findings revealed the existence of a finite temperature difference between the two mirrors when employing this indirect effective coupling approach. This suggests the exciting possibility of engineering a temperature gradient between the mechanical modes using such a setup.
Additionally, we analysed the power spectra of the two mechanical modes and the output spectrum of the second cavity. Remarkably, our study shows that by measuring the latter, it becomes feasible to reconstruct the spectra of the mirrors accurately.
To ensure the stability of the mirror dynamics, we explored the regions of stability as a function of the first cavity pump. Moreover, we delved into the potential existence of multiple steady states for both cavity photon number and mechanical position concerning a specific intensity of light pumped into the first cavity.

The findings regarding correlations, temperature gradients, spectrum reconstruction, stability regions, and multiple steady states open up new possibilities for controlling and manipulating these complex systems for various applications.\\\\

\section*{Acknowledgement}
SL and GMP acknowledge support by MUR under PRIN Project No. 2017 SRN-BRK QUSHIP.

\appendix
\section[\appendixname~\thesection]{}\label{AppA}
\subsection*{Two modes mutual information and quantum discord}
As known by \cite{olivares_quantum_2012} given a two-mode Gaussian state covariance matrix
\begin{equation}
    \mathbf{C}=
    \begin{pmatrix}
        \mathbf{A} & \mathbf{D}\\
        \mathbf{D}^T & \mathbf{B}\\
    \end{pmatrix},
\end{equation}

where  $\mathbf{A}$, $\mathbf{B}$ and $\mathbf{D}$ are $2\times 2$ matrices
one can define four local symplectic invariants, $I_1=det(\mathbf{A})$, $I_2=det(\mathbf{B})$, $I_3=det(\mathbf{D})$, $I_4=det(\mathbf{C})$ and its symplectic eigenvalues 
\begin{equation}
d_{\pm}=\sqrt{\dfrac{I_{\Delta}\pm\sqrt{I_{\Delta}^2-4I_4}}{2}}
\end{equation}
where $I_{\Delta}=I_1+I_2+2I_3$.\\
Mutual information, defined for two quantum systems A and B as 
\begin{equation}
    \mathcal{I}(AB)=S(\rho_A)+S(\rho_B)-S(\rho_{AB})
\end{equation}
(where $S(\rho)$ refers to Von Neumann entropy) can be evaluated, in the case of continous variables gaussian states, in terms of those symplectic invariants as 
\begin{equation}
    \mathcal{I}(AB)=f(\sqrt{I_1})+f(\sqrt{I_2})-f(d_+)-f(d_-)
\end{equation}

with
\begin{equation}
f(x)=\left(x+\dfrac{1}{2}\right)log\left(x+\dfrac{1}{2}\right)-\left(x-\dfrac{1}{2}\right)log\left(x-\dfrac{1}{2}\right).
\end{equation}
Mutual information quantifies the correlations between two quantum systems. Quantum discord, interpreted as the amount on quantumness of these correlations, is then defined as
\begin{equation}
    \mathcal{D}(A|B)=\mathcal{I}(AB)-\mathcal{C}(A|B)
\end{equation}
where 
\begin{equation}
    \mathcal{C}(A|B)=\max_{\{E_a\}}\left[S(\rho_A)-\sum_a p_a S\left(\frac{Tr_B[\rho_{AB}E_a]}{p_a}\right)\right]
\end{equation}
is the total amount of classical correlations and $\sum_a E_a=\mathbf{1}$ is a positive operator value measurement (POVM).\\
In terms of symplectic invariants of covariance matrix, quantum discord can be evaluated as
\begin{equation}
\mathcal{D}(A|B)=f(\sqrt{I_2})+f(d_+)+f(d_-)+f(\sqrt{W}),
\end{equation}
where 
\begin{equation}
    W=
    \begin{cases}
    \left[\frac{2|I_3|+\sqrt{4I^2_3+(4I_2-1)(4I_4-1)}}{(4I_2-1)}\right]^2\text{ if }\frac{4(I_1I_2-I_4)^2}{(I_1+4I_4)(1+4I_2)I^2_3}\leq 1,\\ \\
    \frac{I_1I_2+I_4-I^2_3-\sqrt{(I_1I_2+I_4-I^2_3)^2-4I_1I_2I_4}}{2 I_2}\text{ otherwise.}
    \end{cases}
\end{equation}

\section[\appendixname~\thesection]{}
\subsection*{Bidirectional case}\label{bid}
In order to consider the most general case in which the two subsystems are coupled through a bidirectional (non chiral) waveguide it's necessary to introduce two vacuum input noise operators, $\hat{a}^{in}_R$ and $\hat{a}^{in}_L$, one for each of the direction of propagation in the waveguide, with autocorrelation functions:
\beq
\braket{\hat{a}^{in}_i(t)\hat{a}^{in\dagger}_j(t')}=\delta_{ij}\delta(t-t')\text{\quad with\quad }i,j=R,L 
\eeq
and the same Brownian noise operators $\hat{\xi}_j$ defined in \cref{sec2}.\\
Once defined these new operators, one can straightforwardly follow the procedure described in the previous sections and find the new  set of non linear differential equations for the mean values 
\begin{subequations}
\begin{align}
    &\frac{dQ_j(t)}{dt}=\Om_j P_j(t)\\
    &\frac{dP_j(t)}{dt}=-\Om_jQ_j(t)-\ga_jP_j(t)+|G_j(t)|^2/g_j\\
    &\frac{dA_1(t)}{dt}=-\frac{\kappa}{2}A_1(t)-i\Delta_1(t) A_1(t)-\kappa A_2(t)+E_1\\
    &\frac{dA_2(t)}{dt}=-\frac{\kappa}{2}A_2(t)-i\Delta_2(t)A_2(t)-\kappa A_1(t)+E_2
\end{align}
\end{subequations}
and for the fluctuations one can again find a Lyapunov equation that the covariance matrix $\mathbf{C}$ of the system must obey 
\beq
\frac{d\mathbf C(t)}{dt}=\mathbf S(t)\mathbf C(t)+\mathbf C(t)\mathbf S(t){+}\mathbf N
\eeq
 in which the drift ($\mathbf S$) and diffusion ($\mathbf N$) matrices now become (cfr. \cref{Lyapunov})
\begin{align}
\mathbf S=\begin{pmatrix}
	\mathbf S_1& \mathbf S_{12}\\
	\mathbf S_{12}& \mathbf S_2\end{pmatrix}
\qquad\text{and}\qquad 
	\mathbf N=\begin{pmatrix}
		\mathbf N_1& \mathbf N_{12}\\
		\mathbf N_{12} &\mathbf N_2\end{pmatrix}
\end{align}
with
\begin{align}
\mathbf S_j{=}\begin{pmatrix}
		0&\Om_{j} & 0 & 0 \\
		-\Om_{j}&-\ga & \text{Re}(G_j) & \text{Im}(G_j)\\
		-\text{Im}(G_j) & 0 & -\kappa & \Delta_j\\
		\text{Re}(G_j) & 0 & -\Delta_j & -\kappa
		\end{pmatrix}
\;\;
\mathbf S_{12}{=}\begin{pmatrix}
	0&0 & 0 & 0 \\
	0&0 & 0 & 0 \\
	0&0 & -\kappa & 0 \\
	0&0 & 0 & -\kappa 
\end{pmatrix}
\nonumber
\end{align}
and 
\begin{align}
\mathbf N_j{=}\begin{pmatrix}
	0&0 & 0 & 0 \\
	0&\ga(2 \bar n_j+1) & 0 & 0 \\
	0&0 & 2\kappa & 0 \\
	0&0 & 0 & 2\kappa \\
\end{pmatrix}
\;\;
\mathbf N_{12}{=}\begin{pmatrix}
	0&0 & 0 & 0 \\
	0&0 & 0 & 0 \\
	0&0 & 2\kappa & 0 \\
	0&0 & 0 & 2\kappa
\end{pmatrix}
\nonumber
\end{align}

\begin{figure}[h!]\centering
    \includegraphics[width=.8\columnwidth]{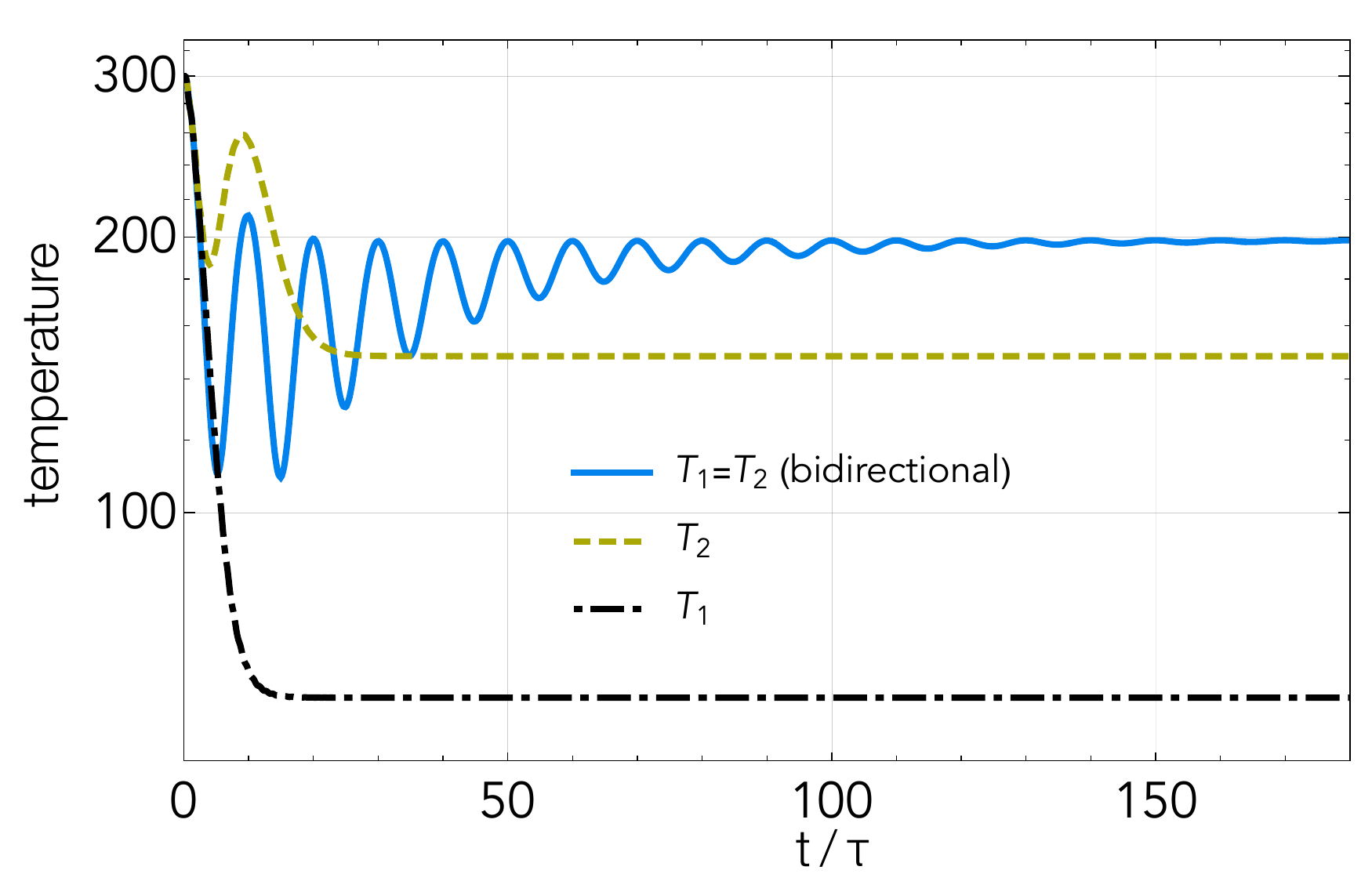}
	\caption{ In this figure is reported the temperature of the two mirrors in the unidirectional (dashed and dot-dashed lines) and bidirectional (solid line) as a function of time, for $E_1=E_2=E$ and $\gamma=2\pi\cdot 10^4$. In order to compare systems with constant total decay rate in the unidirectional case the decay rates were fixed with values $\kappa_R=\kappa$ and $\kappa_L=0$, while in the bidirectional case $\kappa_R=\kappa_L=\kappa/2$. In the first case one can see that a temperature gradient is established while in the latter is not present anymore.}\label{tempbid}
\end{figure}

It's possible now to calculate the temperature as defined in \cref{Temp} and, as shown in \cref{tempbid}, in the case in which the two optomechanical systems are pumped (i.e. $E_2=E_1=E$) it can be seen that no gradient of temperature is established between the two mirrors.

\section[\appendixname~\thesection]{}
\subsection*{Effective Lorenzian peak}
The expression for the spectra in \cref{Sq} can be rearranged, expliciting the expressions of $\chi_{\text{eff}_j}$, $\chi_{j}$ and $\chi_{a_j}$ in the form that, in the limit of $\om_j\simeq\Omega_j$, is a Lorenzian curve. Indeed if we write $\chi_{\text{eff}_j}$ expliciting the susceptibilities we obtain
\begin{align}\label{chieff}
&\chi_j^{\text{eff}}(\omega)=\frac{\chi_j(\om)}{1- i|G_j|^2\chi_j(\om) (\chi_{a_j}(\om)-\chi_{a_j}^*(\minus{\om}))}=\\\nonumber
&=\frac{\Om_j}{\Om_j-\om^2-i\gamma\om- i|G_j|^2\Om_j \left(\frac{1}{\frac{\kappa}{2}-i(\om-\Delta_j)}-\frac{1}{\frac{\kappa}{2}-i(\om+\Delta_j)}\right)}.
\end{align}

\begin{figure}[h!]
	\includegraphics[width=0.9\columnwidth]{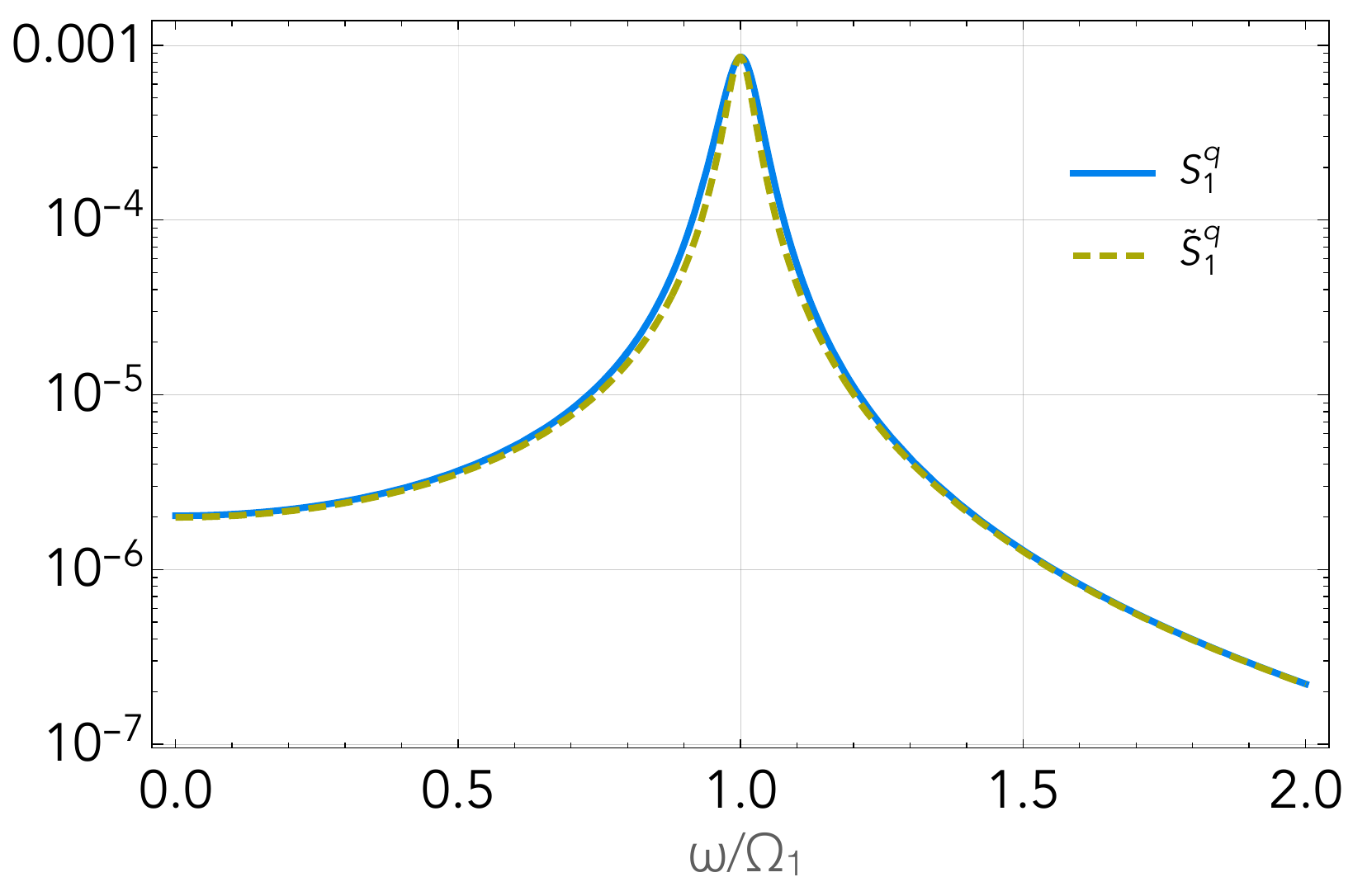}
    \caption{Spectrum of the first mirror plotted using the exact expression     (solid line) and with the one obtained approximating the spectrum as a lorenzian (dashed line)}\label{fit}
\end{figure}

If we define $\tilde\Om_{j}(\om)$ and $\tilde\gamma_{j}(\om)$ as
\begin{align}
&\tilde\Om_{j}(\om)=\sqrt{\Om^2_j-\frac{|G_j|^2 2\Delta_j\Om_j(\frac{\kappa^2}{4}+\Delta^2_j-\om^2)}{(\frac{\kappa^2}{4}+(\om-\Delta_j)^2)(\frac{\kappa^2}{4}+(\om+\Delta_j)^2)}}\\\nonumber
&\tilde\gamma_{j}(\om)=\gamma+\frac{|G_j|^2 2\kappa\Delta_j\Om_j}{(\frac{\kappa^2}{4}+(\om-\Delta_j)^2)(\frac{\kappa^2}{4}+(\om+\Delta_j)^2)}
\end{align}
we can rewrite \cref{chieff} as 
\beq
\chi^{\text{eff}}_{j}(\om)=\frac{\Om_j}{\tilde\Om^2_j(\om)-\om^2-i\tilde\gamma_j(\om)\om}
\eeq
which in the neighborhood of $\om=\Om_j$ can be approximated by
\beq
\chi^{\text{eff}}_{j}(\om)=\frac{\Om_j}{\tilde\Om^2_j(\Om_j)-\om^2-i\tilde\gamma_j(\Om_j)\om}
\eeq
Being proportional to the squared modulus of $\chi^{\text{eff}}_{j}$, the spectra will have the shape of a Lorenzian curve as shown in \cref{fit}.

\section*{References}
\bibliographystyle{iopart-num}
\input{output.bbl}

\providecommand{\newblock}{}

\end{document}

%% file: output.bbl
\providecommand{\newblock}{}

%% file: TempGrad.bbl
\begin{thebibliography}{10}
\expandafter\ifx\csname url\endcsname\relax
  \def\url#1{{\tt #1}}\fi
\expandafter\ifx\csname urlprefix\endcsname\relax\def\urlprefix{URL }\fi
\providecommand{\eprint}[2][]{\url{#2}}

\bibitem{aspelmeyer_quantum_2010}
Aspelmeyer M, Gr{\"o}blacher S, Hammerer K and Kiesel N 2010 {\em Journal of
  the Optical Society of America B\/} {\bf 27} A189 ISSN 0740-3224

\bibitem{bowen_quantum_2020}
Bowen W~P 2020 {\em Quantum Optomechanics / {{Warwick P}}. {{Bowen}}, {{Gerard
  J}}. {{Milburn}}.\/} ({Boca Raton. FL}: {CRC Press, Taylor \& Francis Group})
  ISBN 978-0-367-57519-9

\bibitem{genes_ground-state_2008}
Genes C, Vitali D, Tombesi P, Gigan S and Aspelmeyer M 2008 {\em Physical
  Review A\/} {\bf 77} 033804

\bibitem{gigan_self-cooling_2006}
Gigan S, B{\"o}hm H~R, Paternostro M, Blaser F, Langer G, Hertzberg J~B, Schwab
  K~C, B{\"a}uerle D, Aspelmeyer M and Zeilinger A 2006 {\em Nature\/} {\bf
  444} 67--70 ISSN 14764687

\bibitem{marquardt_quantum_2007}
Marquardt F, Chen J~P, Clerk A~A and Girvin S~M 2007 {\em Physical Review
  Letters\/} {\bf 99} 093902 ISSN 00319007

\bibitem{marquardt_quantum_2008}
Marquardt F, Clerk A~A and Girvin S~M 2008 Quantum theory of optomechanical
  cooling {\em Journal of {{Modern Optics}}\/} vol~55 pp 3329--3338 ISSN
  09500340

\bibitem{yang_ground-state_2019}
Yang J~Y, Wang D~Y, Bai C~H, Guan S~Y, Gao X~Y, Zhu A~D and Wang H~F 2019 {\em
  Optics Express\/} {\bf 27} 22855--22867 ISSN 1094-4087

\bibitem{vinh2021}
Pham V~N~T, Hoang C~M and Vy N~D 2021 {\em Journal of Modern Optics\/} {\bf 68}
  63--71 (\textit{Preprint}
  \eprint{https://doi.org/10.1080/09500340.2021.1875073})
  \urlprefix\url{https://doi.org/10.1080/09500340.2021.1875073}

\bibitem{delic_cooling_2020}
Delić U, Reisenbauer M, Dare K, Grass D, Vuletić V, Kiesel N and Aspelmeyer M
  2020 {\em Science\/} {\bf 367} 892--895 (\textit{Preprint}
  \eprint{https://www.science.org/doi/pdf/10.1126/science.aba3993})
  \urlprefix\url{https://www.science.org/doi/abs/10.1126/science.aba3993}

\bibitem{lecocq_quantum_2015}
Lecocq F, Clark J~B, Simmonds R~W, Aumentado J and Teufel J~D 2015 {\em
  Physical Review X\/} {\bf 5} 041037

\bibitem{pirkkalainen_squeezing_2015}
Pirkkalainen J~M, Damsk{\"a}gg E, Brandt M, Massel F and Sillanp{\"a}{\"a} M~A
  2015 {\em Physical Review Letters\/} {\bf 115} 243601

\bibitem{squeeze_2022}
Magrini L, Camarena-Ch\'avez V~A, Bach C, Johnson A and Aspelmeyer M 2022 {\em
  Phys. Rev. Lett.\/} {\bf 129}(5) 053601
  \urlprefix\url{https://link.aps.org/doi/10.1103/PhysRevLett.129.053601}

\bibitem{palomaki_entangling_2013}
Palomaki T~A, Teufel J~D, Simmonds R~W and Lehnert K~W 2013 {\em Science\/}
  {\bf 342} 710--713

\bibitem{riedinger_non-classical_2016}
Riedinger R, Hong S, Norte R~A, Slater J~A, Shang J, Krause A~G, Anant V,
  Aspelmeyer M and Gr{\"o}blacher S 2016 {\em Nature\/} {\bf 530} 313--316 ISSN
  1476-4687

\bibitem{gut_stationary_2020}
Gut C, Winkler K, {Hoelscher-Obermaier} J, Hofer S~G, Nia R~M, Walk N, Steffens
  A, Eisert J, Wieczorek W, Slater J~A, Aspelmeyer M and Hammerer K 2020 {\em
  Physical Review Research\/} {\bf 2} 033244

\bibitem{Hong}
Hong S, Riedinger R, Marinković I, Wallucks A, Hofer S~G, Norte R~A,
  Aspelmeyer M and Gröblacher S 2017 {\em Science\/} {\bf 358} 203--206
  (\textit{Preprint}
  \eprint{https://www.science.org/doi/pdf/10.1126/science.aan7939})
  \urlprefix\url{https://www.science.org/doi/abs/10.1126/science.aan7939}

\bibitem{bhattacharya_multiple_2008}
Bhattacharya M and Meystre P 2008 {\em Physical Review A\/} {\bf 78} 041801

\bibitem{xuereb_reconfigurable_2014}
Xuereb A, Genes C, Pupillo G, Paternostro M and Dantan A 2014 {\em Physical
  Review Letters\/} {\bf 112} 133604

\bibitem{xuereb_strong_2012}
Xuereb A, Genes C and Dantan A 2012 {\em Physical Review Letters\/} {\bf 109}
  223601

\bibitem{weaver_coherent_2017-1}
Weaver M~J, Buters F, Luna F, Eerkens H, Heeck K, {de Man} S and Bouwmeester D
  2017 {\em Nature Communications\/} {\bf 8} 824 ISSN 2041-1723

\bibitem{piergentili_two-membrane_2021}
Piergentili P, Li W, Natali R, Malossi N, Vitali D and Giuseppe G~D 2021 {\em
  New Journal of Physics\/} {\bf 23} 073013 ISSN 1367-2630

\bibitem{peano_topological_2015}
Peano V, Brendel C, Schmidt M and Marquardt F 2015 {\em Physical Review X\/}
  {\bf 5} 031011

\bibitem{gan_solitons_2016}
Gan J~H, Xiong H, Si L~G, L{\"u} X~Y and Wu Y 2016 {\em Optics Letters\/} {\bf
  41} 2676--2679 ISSN 1539-4794

\bibitem{heinrich_collective_2011}
Heinrich G, Ludwig M, Qian J, Kubala B and Marquardt F 2011 {\em Physical
  Review Letters\/} {\bf 107} 043603

\bibitem{gil-santos_light-mediated_2017}
{Gil-Santos} E, Labousse M, Baker C, Goetschy A, Hease W, Gomez C, Lema{\^i}tre
  A, Leo G, Ciuti C and Favero I 2017 {\em Physical Review Letters\/} {\bf 118}
  063605

\bibitem{pichler_quantum_2015}
Pichler H, Ramos T, Daley A~J and Zoller P 2015 {\em Physical Review A\/} {\bf
  91} 042116

\bibitem{giovannetti_master_2012-3}
Giovannetti V and Palma G~M 2012 {\em Journal of Physics B: Atomic, Molecular
  and Optical Physics\/} {\bf 45} 154003 ISSN 0953-4075

\bibitem{giovannetti_master_2012-2}
Giovannetti V and Palma G~M 2012 {\em Physical Review Letters\/} {\bf 108}
  040401

\bibitem{ramos_quantum_2014}
Ramos T, Pichler H, Daley A~J and Zoller P 2014 {\em Physical Review Letters\/}
  {\bf 113} 237203

\bibitem{cusumano_interferometric_2018-1}
Cusumano S, Mari A and Giovannetti V 2018 {\em Physical Review A\/} {\bf 97}
  053811

\bibitem{karg_remote_2019}
Karg T~M, Gouraud B, Treutlein P and Hammerer K 2019 {\em Physical Review A\/}
  {\bf 99} 063829

\bibitem{Farace2014}
Farace A, Ciccarello F, Fazio R and Giovannetti V 2014 {\em Phys. Rev. A\/}
  {\bf 89}(2) 022335
  \urlprefix\url{https://link.aps.org/doi/10.1103/PhysRevA.89.022335}

\bibitem{li_long-distance_2016}
Li T, Bao T~Y, Zhang Y~L, Zou C~L, Zou X~B and Guo G~C 2016 {\em Optics
  Express\/} {\bf 24} 12336--12348 ISSN 1094-4087 publisher: Optical Society of
  America
  \urlprefix\url{https://www.osapublishing.org/oe/abstract.cfm?uri=oe-24-11-12336}

\bibitem{Li_Jin}
Li J, Zhou Z~H, Wan S, Zhang Y~L, Shen Z, Li M, Zou C~L, Guo G~C and Dong C~H
  2022 {\em Phys. Rev. Lett.\/} {\bf 129}(6) 063605
  \urlprefix\url{https://link.aps.org/doi/10.1103/PhysRevLett.129.063605}

\bibitem{xuereb_routing}
Xuereb A, Barzanjeh S and Aquilina M 2018 Routing thermal noise through quantum
  networks p~59

\bibitem{law_interaction_1995}
Law C~K 1995 {\em Physical Review A\/} {\bf 51} 2537--2541 ISSN 10502947

\bibitem{genes_chapter_2009}
Genes C, Mari A, Vitali D and Tombesi P 2009 Chapter 2 {{Quantum Effects}} in
  {{Optomechanical Systems}}

\bibitem{giovannetti_phase-noise_2001}
Giovannetti V and Vitali D 2001 {\em Physical Review A - Atomic, Molecular, and
  Optical Physics\/} {\bf 63} 1--8 ISSN 10502947

\bibitem{gardiner_driving_1994}
Gardiner C~W and Parkins A~S 1994 {\em Physical Review A\/} {\bf 50} 1792--1806
  ISSN 10502947

\bibitem{gardiner_quantum_nodate}
Gardiner C~W and Zoller P {\em Quantum {{Noise}}\/}

\bibitem{gardiner_input_1985}
Gardiner C~W and Collett M~J 1985 {\em Physical Review A\/} {\bf 31} 3761--3774
  ISSN 10502947

\bibitem{wilson-rae_theory_2007}
{Wilson-Rae} I, Nooshi N, Zwerger W and Kippenberg T~J 2007 {\em Physical
  Review Letters\/} {\bf 99} 093901 ISSN 00319007

\bibitem{Serafini}
Serafini G, Zippilli S and Marzoli I 2020 {\em Phys. Rev. A\/} {\bf 102}(5)
  053502 \urlprefix\url{https://link.aps.org/doi/10.1103/PhysRevA.102.053502}

\bibitem{olivares_quantum_2012}
Olivares S 2012 {\em European Physical Journal: Special Topics\/} {\bf 203}
  3--24 ISSN 19516355

\bibitem{Zurek2001}
Ollivier H and Zurek W~H 2001 {\em Phys. Rev. Lett.\/} {\bf 88}(1) 017901
  \urlprefix\url{https://link.aps.org/doi/10.1103/PhysRevLett.88.017901}

\bibitem{Adesso2010}
Adesso G and Datta A 2010 {\em Phys. Rev. Lett.\/} {\bf 105}(3) 030501
  \urlprefix\url{https://link.aps.org/doi/10.1103/PhysRevLett.105.030501}

\bibitem{Giorda2010}
Giorda P and Paris M~G~A 2010 {\em Phys. Rev. Lett.\/} {\bf 105}(2) 020503
  \urlprefix\url{https://link.aps.org/doi/10.1103/PhysRevLett.105.020503}

\bibitem{paternostro_reconstructing_2006}
Paternostro M, Gigan S, Kim M~S, Blaser F, B{\"o}hm H~R and Aspelmeyer M 2006
  {\em New Journal of Physics\/} {\bf 8} 107--107 ISSN 13672630

\bibitem{ludwig_optomechanical_2008}
Ludwig M, Kubala B and Marquardt F 2008 {\em New Journal of Physics\/} {\bf 10}
  ISSN 13672630

\bibitem{marquardt_dynamical_2006}
Marquardt F, Harris J~G~E and Girvin S~M 2006 {\em Physical Review Letters\/}
  {\bf 96} 103901

\bibitem{Sarma:16}
Sarma B and Sarma A~K 2016 {\em J. Opt. Soc. Am. B\/} {\bf 33} 1335--1340
  \urlprefix\url{https://opg.optica.org/josab/abstract.cfm?URI=josab-33-7-1335}

\end{thebibliography}
